\documentclass{article}
\usepackage{spconf}

\usepackage[]{siunitx}

\usepackage{cite}
\usepackage{graphicx}
\usepackage[cmex10]{amsmath}
\usepackage{multirow}
\usepackage{makecell}

\usepackage{newtxtext,newtxmath}
\usepackage{bm}
\usepackage{amssymb}
\usepackage{psfrag}
\usepackage{color}
\usepackage{colortbl}

\def\argmax{\mathop{\rm argmax}}

\title{Advances in Online Audio-Visual Meeting Transcription}
\makeatletter
\def\@name{\textit{Takuya Yoshioka, Igor Abramovski, Cem Aksoylar, Zhuo Chen, Moshe David, Dimitrios Dimitriadis}\\
\textit{Yifan Gong, Ilya Gurvich, Xuedong Huang, Yan Huang, Aviv Hurvitz, Li Jiang, Sharon Koubi}\\
\textit{Eyal Krupka, Ido Leichter, Changliang Liu, Partha Parthasarathy, Alon Vinnikov, Lingfeng Wu}\\
\textit{Xiong Xiao, Wayne Xiong, Huaming Wang, Zhenghao Wang, Jun Zhang, Yong Zhao, Tianyan Zhou}\\
}
\makeatother
\address{\vspace{-.5em}Microsoft, One Microsoft Way, Redmond, WA, USA}

\begin{document}
\ninept

\maketitle

\begin{abstract}
This paper describes a system that generates speaker-annotated transcripts of meetings by using a microphone array and a 360-degree camera. 
The hallmark of the system is its ability to handle overlapped speech, which 
has been an unsolved problem in realistic settings for over a decade. 
We show that this problem can be addressed by using a continuous speech separation approach. 
In addition, we describe an online audio-visual speaker diarization method 
that leverages face tracking and identification, sound source localization, speaker identification, and, if available, prior speaker information for robustness 
to various real world challenges. 
All components are integrated in a meeting transcription framework called SRD, which stands for ``separate, recognize, and diarize''.
Experimental results using recordings of natural meetings involving up to 11 attendees are reported. 
The continuous speech separation improves a word error rate (WER) by 16.1\%
compared with a highly tuned beamformer. 
When a complete list of meeting attendees is available, 
the discrepancy between WER and speaker-attributed WER is only 1.0\%, indicating accurate word-to-speaker association. 
This increases marginally to 1.6\% when 50\% of the attendees are unknown to the system. 
\end{abstract}
\section{Introduction}
\label{sec:Introduction}
\vspace{-.5em}

The goal of meeting transcription is to have machines generate
speaker-annotated transcripts of natural meetings based on their audio and optionally video recordings. 
Meeting transcription and analytics would be a key to enhancing productivity as well as improving accessibility in the workplace. 
It can also be used for conversation transcription in other domains such as 
healthcare~\cite{Chiu18}. 
Research in this space was promoted in the 2000s by NIST Rich Transcription Evaluation series and public release of relevant corpora~\cite{FiscusEtAl:rt07,Janin03,Carletta06}. 
While systems developed in the early days yielded high error rates, 
advances have been made in individual component technology fields, including conversational 
speech recognition~\cite{Xiong16,Saon17}, far-field speech processing~\cite{Yoshioka15b,Du16,Li17}, and speaker identification and diarization~\cite{Dimitriadis17,Zhang18,Sell18}.
When cameras are used in addition to microphones to capture the meeting conversations, 
speaker identification quality could be further improved
thanks to the computer vision technology. 
These trends motivated us to build an end-to-end audio-visual meeting transcription system to identify and address unsolved challenges. 
This report describes our learning, with focuses on 
overall architecture design, overlapped speech recognition, and audio-visual speaker diarization. 

When designing meeting transcription systems, different constraints must be taken into account depending on targeted scenarios. 
In some cases, microphone arrays are used as an input device. 
If the names of expected meeting attendees are known beforehand, the transcription system should be able to provide each utterance with the true identity (e.g., ``Alice'' or ``Bob'') instead of a randomly generated label like ``Speaker1''. 
It is often required to show the transcription in near real time, which makes the task more challenging. 

This work assumes the following scenario. 
We consider a scheduled meeting setting, where an organizer arranges a meeting in advance and sends invitations to attendees. The transcription system has access to the invitees' names. However, actual attendees may not completely match those invited to the meeting. The users are supposed to enroll themselves in the system beforehand so that their utterances in the meeting can be associated with their names. 
The meeting is recorded with an audio-visual device equipped with a seven-element circular microphone array and a fisheye camera. Transcriptions must be shown with a latency of up to a few seconds. 

This paper investigates three key challenges.

\vspace{.2em}
\noindent
\textbf{Speech overlaps:}~~Recognizing overlapped speech has been one of the main challenges in meeting transcription with limited tangible progress. 
Numerous multi-channel speech separation methods were proposed based on independent component analysis or spatial clustering~\cite{Sawada04,Buchner05,Nesta11,Sawada11,Ito14,Drude17}. However, 
there was little successful effort to apply these methods to natural meetings.  Neural network-based single-channel separation methods using techniques like 
permutation invariant training (PIT)~\cite{Kolbaek17} or deep clustering (DC)~\cite{Hershey16} are known to be vulnerable to various types of acoustic distortion, including reverberation and background noise~\cite{Maciejewski18}. 
In addition, these methods were tested almost exclusively on small-scale segmented synthetic data and have not been applied to continuous conversational speech audio. 
Although the recently held CHiME-5 challenge helped the community make a step forward 
to a realistic setting, it still allowed the use of ground-truth speaker segments~\cite{Barker18,Kanda19}.

We address this long-standing problem with a continuous speech separation (CSS) approach, which we proposed in our latest conference papers~\cite{Yoshioka18b,Yoshioka19}. 
It is based on an observation that 
the maximum number of simultaneously active speakers is usually limited even in a large meeting. 
According to \cite{Cetin06}, two or fewer speakers are active for more than 98\% of the meeting time. 
Thus, given continuous multi-channel audio observation, we generate a fixed number, say $N$, of
time-synchronous signals. Each utterance is separated from overlapping voices and background noise.
Then, the separated utterance is spawned from one of the $N$ output
channels. 
For periods where the number of active speakers is fewer than $N$, the extra channels generate zeros. 
We show how continuous speech separation can fit in with an overall meeting transcription architecture to generate speaker-annotated transcripts. 

Note that our speech separation system does not make use of a camera signal. While much progress has been made in audio-visual speech separation, the challenge of dealing with all kinds of image variations remains unsolved~\cite{Ephrat18,Afouras18,Wu19}.

\vspace{.3em}
\noindent
\textbf{Extensible framework:}~~It is desirable that a single transcription system be able to support various application settings for both maintenance and scalability purposes.
While this report focuses on the audio-visual setting, 
our broader work covers an audio-only setting as well as the scenario where no prior knowledge of meeting attendees is available. 
A modular and versatile architecture is desired to encompass these different settings.

To this end, we propose a framework called SRD, which stands for ``separate, recognize, and diarize'', where CSS, 
speech recognition, and speaker diarization takes place in tandem. 
Performing CSS at the beginning allows the other modules to operate on overlap-free signals.
Diarization is carried out after speech recognition because its implementation can vary significantly depending on the application settings. 
By choosing an appropriate diarization module for each setting, multiple use cases can be supported without changing the rest of the system.  
This architecture also allows transcriptions to be displayed in real time without speaker information. 
Speaker identities for each utterance may be shown after a couple of seconds.


\vspace{.3em}
\noindent
\textbf{Audio-visual speaker diarization:}~~Speaker diarization, a process of segmenting input audio and assigning speaker labels to the individual segments, 
can benefit from a camera signal. 
The phenomenal improvements that have been made to face detection and identification algorithms by convolutional neural networks (CNNs)~\cite{liu2017sphereface, taigman2014deepface, HeGDG17} make the camera signal very appealing for speaker diarization.
While much prior work assumes the batch processing scenario where
the entire meeting recording can be processed multiple times, 
several studies deal with online processing~\cite{Schmalenstroeer10,Hori12,Gebru18,Ban18}. 
However, no previous studies comprehensively address the challenges 
that one might encounter in real meetings. 
\cite{Schmalenstroeer10,Hori12} do not cope with speech overlaps. 
While the methods proposed in \cite{Gebru18,Ban18} address the overlap issue, they rely solely on spatial cues and thus are not applicable when multiple speakers sit side by side. 


Our diarization method handles overlapping utterances as well as co-located speakers 
by utilizing the time-frequency (TF) masks generated by CSS in speaker identification and sound source localization (SSL).
In addition, several enhancements are made to face identification to improve robustness
to image variations caused by face occlusions, extreme head pose, lighting conditions, and so on. 


\section{Device and Data}
\label{sec: device}
\vspace{-.5em}

Our audio-visual diarization approach leverages spatial information and thus requires the audio and video angles to align. Because existing meeting corpora do not meet this requirement, we collected audio-visual English meeting recordings at Microsoft Speech and Language Group with 
an experimental recording device. 

Our device has a cone shape and is approximately 30 centimeters high, slightly higher than a typical laptop. 
At the top of the device is a fisheye camera, providing a 360-degree field of view. 
Around the middle of the device, there is a horizontal seven-channel circular microphone array. 
The first microphone is placed at the center of the array board while the other microphones 
are arranged along the perimeter with an equal angle spacing. 
The board is about 10 cm wide.  

The meetings were recorded in various conference rooms. 
The recording device was placed at a random position on a table in each room. 
We had meeting attendees sign up for the data collection program and go through audio and video enrollment steps. For each attendee, we obtained approximately a voice recording of 20 to 30 seconds and 10 or fewer close-up photos from different angles. 
A total of 26 meetings were recorded for the evaluation purpose. 
Each meeting had a different number of attendees, ranging from 2 to 11. 
The total number of unique participants were 62. 
No constraint was imposed on seating arrangements. 

Two test sets were created: a gold standard test set and an extended test set. They were manually transcribed in different ways. 
The gold standard test set consisted of seven meetings and was 4.0 hours long in total. 
Those meetings were recorded both with the device described above and headset microphones. 
Professional transcribers were asked to provide initial transcriptions
by using the headset and far-field audio recordings as well as the video. Then, automatic segmentation was performed with forced alignment. Finally, the segment boundaries and transcriptions were reviewed and corrected. 
Significant effort was made to fine-tune timestamps of the segmentation boundaries. 
While being very accurate, this transcription process requires headset recordings and therefore is not scalable. 
The extended test set contained 19 meetings totaling 6.4 hours. It covered a wider variety of conditions. 
These additional meetings were recorded only with the audio-visual device, i.e., 
the participants were not tethered to headsets. 
In addition to the audio-visual recordings, the transcribers were provided with outputs of our prototype system to bootstrap the transcription process.

\section{Separate-Recognize-Diarize Framework}
\label{sec: architecture}
\vspace{-.5em}

\begin{figure*}
    \centering
    \includegraphics[scale=0.64]{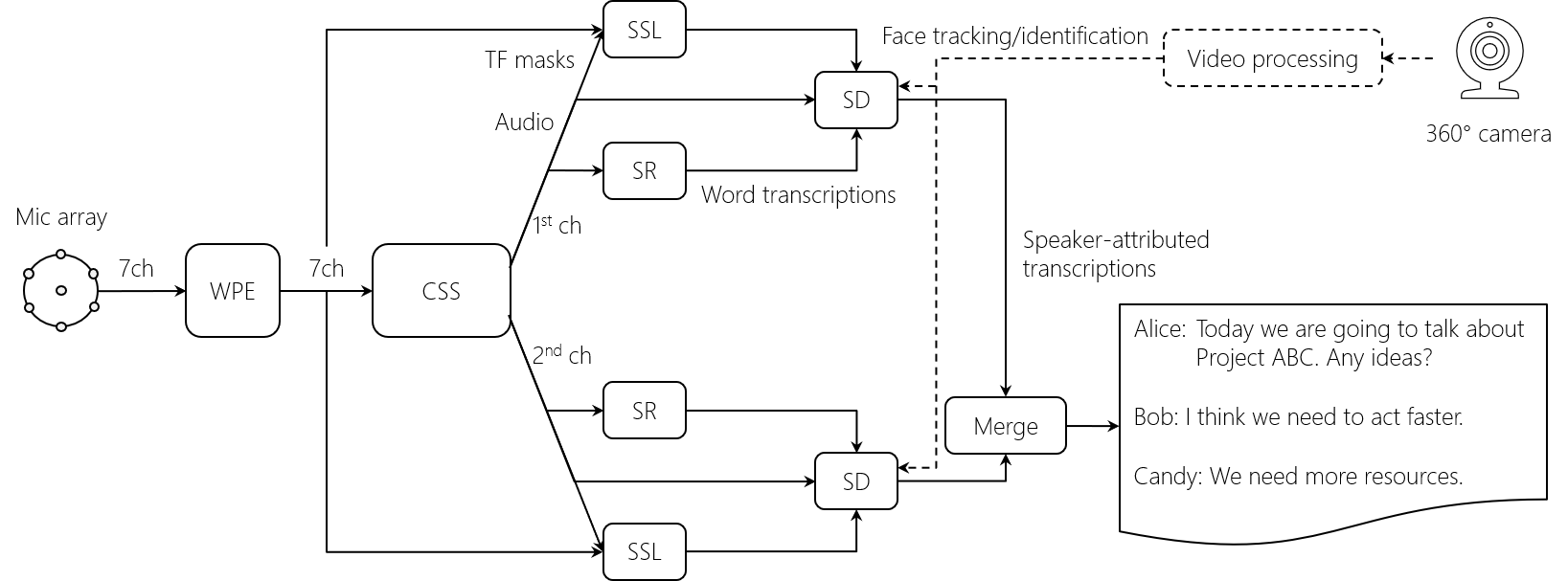}
    \caption{Processing flow diagram of SRD framework for two stream configuration. To run the whole system online, the video processing and SR modules are assigned their own dedicated resources. 
    WPE: weighted prediction error minimization for dereverberation. CSS: continuous speech separation. SR: speech recognition. SD: speaker diarization. SSL: sound source localization.}
    \label{fig: SRD}
    \vspace{-.5em}
\end{figure*}

Figure~\ref{fig: SRD} shows a processing flow of the SRD framework for generating speaker-annotated transcripts.  
First, multi-input multi-output dereverberation is performed in real time~\cite{Yoshioka12c}. 
This is followed by CSS, which generates $N$ distinct signals (the diagram shows the case of $N$ being 2). 
Each signal has little overlapped speech, which allows for the use of 
conventional speech recognition and speaker diarization modules. 
After CSS, speech recognition is performed using each separated signal. 
This generates a sequence of speech events, where each event consists of a sequence of time-marked recognized words. 
The generated speech events are fed to a speaker diarization module to label each recognized word 
with the corresponding speaker identity.
The speaker labels may be taken from a meeting invitee list or automatically generated by the system, like "Speaker1". 
Finally, the speaker-annotated transcriptions from the $N$ streams are merged. 




\vspace{.2em}
\noindent
\textbf{Comparison with other architectures:}~~Most prior work in multi-microphone-based meeting transcription performs acoustic beamforming to generate a single enhanced audio signal, which is then processed with speaker diarization and speech recognition~\cite{Hain12}. 
This scheme fails in transcription in overlapped regions which typically make up more than 10\% of the speech period. 
It is also noteworthy that 
beamforming and speaker diarization tend to suffer if speakers exchange turns quickly one after another even when their utterances do not overlap. 

The system presented in \cite{Hori12} uses speaker-attributed beamforming, which generates a separate signal for each speaker. The speaker-attributed signals are processed with speech recognition to generate transcriptions for each speaker. This requires accurate speaker diarization to be performed in real time before beamforming, which is challenging in natural meetings. 

By contrast, by performing CSS at the beginning, the SRD approach can handle overlaps of up to $N$ speakers
without special overlap handling in speech recognition or speaker diarization. 
We also found that performing diarization after speech recognition resulted in more accurate transcriptions than the conventional way of performing diarization before speech recognition. One reason is that the ``post-SR'' diarization can take advantage of the improved speech activity detection capability offered by the speech recognition module. 
Also, the speaker change positions can be restricted to word boundaries. 
The same observation was reported in \cite{Dimitriadis17}.

\section{Continuous Speech Separation}
\label{sec: separation}
\vspace{-.5em}

The objective of CSS is to render an input multi-channel signal containing overlaps into multiple overlap-free signals. 
Conceptually, CSS monitors the input audio stream; when overlapping utterances are found, it isolates these utterances and distributes them to different output channels. 
Non-overlapped utterances can be output from one of the channels. 
We want to achieve this in a streaming fashion without explicitly 
performing segmentation or overlap detection. 

\begin{figure}[t]
    \centering
    \includegraphics[scale=0.55]{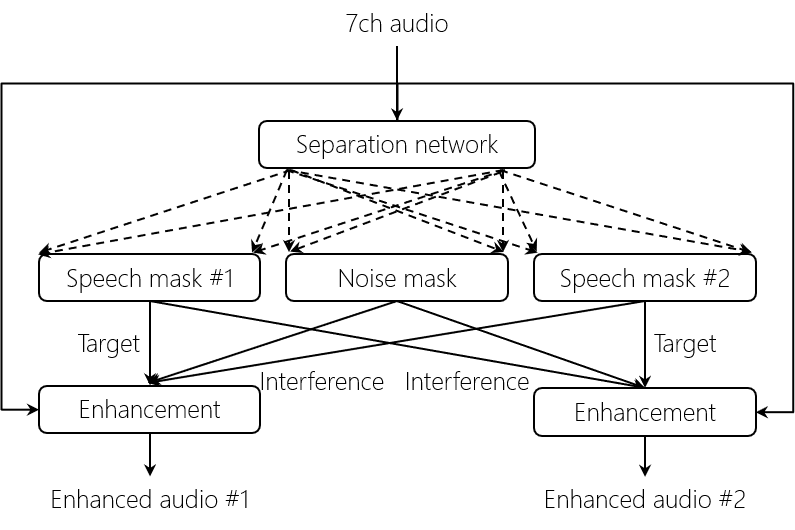}
    \vspace{-1.5em}
    \caption{Speech separation processing flow diagram.}
    \label{fig: CSS}
    \vspace{-.5em}
\end{figure}

We perform CSS by using a speech separation network trained with PIT as we first proposed in \cite{Yoshioka18b}.
Figure~\ref{fig: CSS} shows our proposed CSS processing flow for the case of $N=2$.
First, single- and multi-channel features are extracted for each short time frame from an input seven-channel signal. 
The short time magnitude spectral coefficients of the center microphone and the inter-channel phase differences (IPDs) with reference to the center microphone are used as the single- and multi-channel features, respectively. 
The features are mean-normalized with a sliding window of four seconds and then fed to a speech separation network, which yields $N$ different speech masks as well as a noise mask for each TF bin. 
A bidirectional long short time memory (BLSTM) network is employed to leverage
long term acoustic dependency. 
Finally, for each $n \in \{0, \cdots, N-1\}$,  
the $n$th separated speech signal is generated by enhancing the speech component articulated by the $n$th speech TF masks while suppressing those represented by the other masks. 
To generate the TF masks in a streaming fashion with the bidirectional model, this is repeated every 0.8 seconds by using a 2.4-second segment. 
It should be noted that the speech separation network may change the order of the $N$ speech outputs when processing different data segments. In order to align the output order of the current segment with that 
of the previous segment, the best order is estimated by examining all possible permutations. 
The degree of ``goodness'' of each permutation is measured as the mean squared error 
between the masked magnitude spectrograms calculated over the frames shared by the two adjacent segments.



Given the $N+1$ TF masks ($N$ for speech, one for noise),  
we generate each of the $N$ output signals with mask-based minimum variance distortionless response (MVDR) beamforming\footnote{Fixed beamformers may be used instead, together with post-processing neural networks~\cite{Chen18}.}~\cite{Yoshioka18b}.
The MVDR filter for each output channel is updated periodically, every 0.8 seconds in our implementation. 
We follow the MVDR formula of equation (24) of \cite{Souden10}.
This scheme requires the spatial covariance matrices (SCMs) of the target and interference signals, where the interference signal means the sum of all non-target speakers and the background noise. 
To estimate these statistics, 
we continuously estimate the target SCMs for all the output channels as well as the noise SCM, with a refresh rate of 0.8 seconds. 
The noise SCM is computed by using a long window of 10 seconds, considering the fact that the background noise tends to be stationary in conference rooms. 
On the other hand, the target SCMs are computed with a relatively short window of 2.4 seconds. 
The interference SCM for the $n$th output channel is then obtained by adding up the noise SCM and all the target SCMs except that of the $n$th channel. 

\vspace{.2em}
\noindent
\textbf{Separation model details:}~~Our speech separation model is comprised of a three-layer 1024-unit BLSTM. 
The input features are transformed by a 1024-unit
projection layer with ReLU nonlinearity before being fed to the
BLSTM. On top of the last BLSTM layer, there is a three-headed
fully connected sigmoid layer assuming $N$to be 2, where each head produces
TF masks for either speech or noise. 

The model is trained on 567 hours of artificially generated noisy and reverberant speech mixtures. 
Source speech signals are taken from WSJ SI-284 and LibriSpeech.
Each training sample is created as follows. First,
the number of speakers (1 or 2) is randomly chosen. For the
two-speaker case, the start and end times of each utterance is
randomly determined so that we have a balanced combination of the
four mixing configurations described in \cite{Yoshioka18}. The source signals are
reverberated with the image method~\cite{Allen79}, mixed together in the
two-speaker case, and corrupted by additive noise. The multi-channel
additive noise signals are simulated by assuming a
spherically isotropic noise field. Long training samples
are clipped to 10 seconds. The model is trained to minimize the PIT-MSE between the source magnitude spectra and the masked versions of the observed magnitude spectra. As noted in \cite{Yoshioka18b}, PIT is applied only to the two speech masks.

\section{Speaker Diarization}
\label{sec: diarization}
\vspace{-.5em}

Following the SRD framework, 
each CSS output signal is processed with speech recognition and then speaker diarization. 
The input to speaker diarization is a speech event, a sequence of recognized words between silent periods in addition to the audio and video signals of the corresponding time segment.  
The speaker diarization module attributes each word to the person who is supposed to have spoken that word. 
Note that speaker diarization often refers to a process of assigning anonymous (or relative~\cite{Tranter06}) speaker labels~\cite{Anguera12}. 
Here, we use this term in a broader way: we use true identities, i.e., real names, when they are invited through the conferencing system.

Speaker diarization is often performed in two steps: segmentation and speaker attribution. 
The segmentation step decomposes the received speech event into speaker-homogeneous subsegments. 
Preliminary experiments showed that 
our system was not very sensitive to the choice of a segmentation method\footnote{The relative SA-WER difference between the  HMM segmentation and the agglomerative clustering-based segmentation used in \cite{Yoshioka19b} was less than 1\%.}.
This is because, even when two persons speak one after the other, their signals are likely to be assigned to different CSS output channels~\cite{Yoshioka18}.
In other words, CSS undertakes the segmentation to some extent. 
Therefore, in this paper, we simply use 
a hidden Markov model-based method that is similar to the one proposed in \cite{Schmalenstroeer10}. 

The speaker attribution step finds the most probable speaker ID for a given segment by using the audio and video signals. This is formalized as
\begin{align}
\Hat{h} = \argmax_{h \in \mathcal{H}} P(h | A, V; M). \label{eq: map}
\end{align}
$A$ and $V$ are the audio and video signals, respectively. $M$ is the set of the TF masks of the current CSS channel within the input segment. 
The speaker ID inventory, $\mathcal{H}$, consists of the invited speaker names (e.g., `Alice' or `Bob') and anonymous `guest'  IDs produced by the vision module (e.g., `Speaker1' or `Speaker2')\footnote{In our implementation, $\mathcal{H}$ also contains a special tag which collectively represents speakers 
who have been neither invited to the meeting nor identified by the vision module as guests. 
This can kick in when the vision module failed to report the presence of a guest speaker. Because the impact that this tag has on the overall performance is marginal, we omit the description.}. 
In what follows, we propose a model for combining 
face tracking, face identification, speaker identification, SSL, and the TF masks generated by the preceding CSS module to calculate the speaker ID posterior probability of equation \eqref{eq: map}. 
The integration of these complementary cues would make speaker attribution robust to real world challenges, including speech overlaps, speaker co-location, and the presence of guest speakers. 

First, by treating the face position trajectory of the speaking person as a latent variable,
the speaker ID posterior probability can be represented as 
\begin{align}
P(h | A, V; M) = \sum_{r \in \mathcal{R}} P(h, r | A, V; M), \label{eq: temp0}
\end{align}
where $\mathcal{R}$ includes all face position trajectories detected by the face tracking module within the input period. 
We call a face position trajectory a tracklet. 
The joint posterior probability on the right hand side (RHS) can be factorized as 
\begin{align}
P(h, r | A, V; M) = P(h | r, A, V; M) P(r | A, V; M). \label{eq: temp1}
\end{align}

The RHS first term, or the tracklet-conditioned speaker ID posterior, can be further decomposed as 
\begin{align}
P(h | r, A, V; M) &\propto  P(h | r, V; M) p(A | h, r, V; M) \notag \\
&=P(h | r, V) p(A | h; M).
\end{align}
The RHS first term, calculating the speaker ID posterior given the video signal and the tracklet calls for a face identification model because 
the video signal and the tracklet combine to specify a single speaker's face. 
On the other hand, the likelihood term on the RHS can be calculated as 
\begin{align}
p(A | h; M) = p(A_s | h; M) p(A_m | h; M), 
\end{align}
where we have assumed the spatial and magnitude features of the audio, represented as $A_s$ and $A_m$, respectively, to be independent of each other. 
The RHS first term, $p(A_s | h; M)$, is a spatial speaker model, measuring 
the likelihood of speaker $h$ being active given spatial features $A_s$.
We make no assumption on the speaker positions. Hence, 
$p(A_s | h; M)$ is constant and can be ignored. 
The RHS second term, $p(A_m | h; M)$, is a generative model for speaker identification. 

Returning to \eqref{eq: temp1}, the RHS second term, describing the probability of 
the speaking person's face being $r$ (recall that each tracklet captures a single person's face), may be factorized as 
\begin{align}
P(r | A, V; M) = p(A_s | r; M) P(r | V, A_m; M). 
\end{align}
The first term is the likelihood of tracklet $r$ generating a sound with spatial features $A_s$ and therefore related to SSL. 
The second term is the probability with which the tracklet $r$ is active given the audio magnitude features and the video. Calculating this requires lip sync to be performed for each tracklet, 
which is hard in our application due to low resolution resulting from speaker-to-camera distances and compression artifacts. 
Thus, we ignore this term. 

Putting the above equations together, the speaker-tracklet joint posterior needed in \eqref{eq: temp0} can be obtained as 
\begin{align}
P(h, r | A, V; M) = P(h|r, V) p(A_m | h; M) p(A_s | r; M),
\end{align}
where the ingredients of the RHS relate to 
face identification, speaker identification, and SSL, respectively, 
in the order of appearance. 
The rest of this section describes our implementations of these models. 

\subsection{Sound source localization}
\vspace{-.5em}

The SSL generative model, $p(A_s | r; M)$, is defined by using a complex angular central Gaussian model (CACGM)~\cite{Ito16}. 
The SSL generative model can be written as follows: 
\begin{align}
p(A_s | r; M) = \sum_{\omega} P(\omega | r) p(A_s | \omega; M), 
\end{align}
where $\omega$ is a discrete-valued latent variable representing the sound direction. 
It should be noted that the strongest sound direction may be mismatched with the face direction to a varying degree due to sound reflections on tables, diffraction on obstacles, face orientation variability, and so on. 
$P(\omega | r)$ is introduced to represent this mismatch and modeled as a uniform distribution with a width of 25 degrees centered at the face position for $r$.
The likelihood term, $p(A_s | \omega; M)$, is modeled with the CACGM and the log likelihood reduces to the following form~\cite{Yoshioka19}: 
$
\log p(A_s | \omega ;M) = -\sum_{t,f} m_{t,f} \log (1 - || \bm{z}_{t,f}^H \bm{h}_{f,\omega} ||^2 / (1 + \epsilon) ), 
$
where $\bm{z}_{t,f}$ is a magnitude-normalized multi-channel observation vector constituting $A_s$, $m_{t,f}$ a TF mask, $\bm{h}_{f, \omega}$ a steering vector corresponding to sound direction $\omega$, and $\epsilon$ a small flooring constant. 

\subsection{Speaker identification}
\vspace{-.5em}

As regards the speaker identification model, $p(A_m | h; M)$, 
we squash the observations to a fixed-dimensional representation, i.e. speaker embedding. The proximity in the embedding space measures the similarity between speakers.

Our model consists of multiple convolutional layers augmented by residual blocks~\cite{He16} and has a bottleneck layer. The model is trained to reduce classification errors for a set of known identities. For inference, the output layer of the model is removed and the activation of the bottleneck layer is extracted as a speaker embedding, which is expected to generalize to any speakers beyond those included in the training set. 
In our system, the speaker embedding has 128 dimensions. VoxCeleb corpus~\cite{vox1,vox2} is used for training. Our system was confirmed to outperform the state-of-the-art on the VoxCeleb test set.

We assume an embedding vector of each speaker to follow a von Mises-Fisher distribution with a shared concentration parameter. If we ignore a constant term, this leads to the following equation: 
$\log p(A_m | h; M) = \bm{p}_h^T \bm{d}_M$, 
where $\bm{d}_M$ is the embedding extracted from the signal enhanced with the TF masks in $M$, and $\bm{p}_h$ is speaker $h$'s mean direction in the embedding space. 
This is equivalent to measuring the proximity of the input audio segment to speaker $h$ by using a cosine similarity in the embedding space~\cite{Banerjee05}. 

The mean direction of a speaker can be regarded as a voice signature of that person. It is calculated as follows. 
When speaker $h$ is an invited speaker, the system has the enrollment audio of this person. Embedding vectors are extracted from the enrollment sound with a sliding window and averaged to produce the mean direction vector. 
For a guest speaker detected by the vision module, no enrollment audio is available at the beginning. The speaker log likelihood, $\log p (A_m | h; M)$, is assumed to have a constant value which is determined by a separate speaker verification experiment on a development set. 
For both cases, $\bm{p}_h$, the voice signature of speaker $h$, is updated during the meeting every time a new segment is attributed to that person.

\subsection{Face tracking and identification}
\vspace{-.5em}

Our vision processing module (see Fig. \ref{fig: SRD}) locates and identifies all persons in a room for each frame captured by the camera. 
The unconstrained meeting scenario involves many challenges, including face occlusions, extreme head pose, lighting conditions, compression artifacts, low resolution due to device-to-person distances, motion blur. Therefore, any individual frame may not contain necessary information. For example, a face may not be detectable in some frames. Even if it is detectable, it may not be recognizable.

To handle this variability, we integrate information across time using face tracking as implied by our formulation of $P(h | r, V)$, which requires face identification to be performed only at a tracklet level. Our face tracking uses face detection and low-level tracking to maintain a set of tracklets, 
where each tracklet is defined as 
a sequence of faces in time that belong to the same person.
We use a method similar to that in \cite{Ren08} with several adaptions to our specific setting, such as exploiting the stationarity of the camera for detecting motion, performing the low-level tracking by color based mean-shift instead of gray-level based normalized correlation, tuning the algorithm to minimize the risk of tracklet mergers (which in our context are destructive), etc. 
Also, the faces in each tracklet are augmented with attributes such as face position, dimensions, head pose, and face feature vectors. 
The tracklet set defines $\mathcal{R}$ of equation \eqref{eq: temp0}\footnote{It is also sensible to add a special tracklet tag representing failure in detecting an active speaker, which we ignore in this paper.}. 

Face identification calculates person ID posterior probabilities for each tracklet. 
Guest IDs (e.g., 'Speaker1') are produced online, each representing a unique person in the meeting who is not on the invitee list. We utilize a discriminative face embedding which converts face images into fixed-dimensional feature vectors, or 128-dimensional vectors obtained as 
output layer activations of a convolutional neural network. 
For the face embedding and detection components, we use the algorithms from Microsoft Cognitive Services Face API~\cite{ming2019group,chen2014joint}. Face identification of a tracklet is performed by comparing the set of face features extracted from its face instances, to the set of features from a gallery of each person's faces. For invited people, the galleries are taken from their enrollment videos, while for guests, the gallery pictures are accumulated online from the meeting video. We next describe our set-to-set similarity measure designed to perform this comparison.

Our set-to-set similarity is designed to utilize information from multiple frames while remaining robust to head pose, lighting conditions, blur and other misleading factors. We follow the matched background similarity (MBGS) approach of~\cite{wolf2011face} and make crucial adaptations to it that increase accuracy significantly for our problem. 
As with MBGS, we train a discriminative classifier for each identity $h$ in $\mathcal{H}$. The gallery of $h$ is used as positive examples, while a separate fixed background set $B$ is used as negative examples. This approach has two important benefits. First, it allows us to train a classifier adapted to a specific person.
Second, the use of a background set $B$ lets us account for misleading sources of variation e.g. if a blurry or poorly lit face from $B$ is similar to one of the positive examples, the classifier's decision boundary can be chosen accordingly.
During meeting initialization, an support vector machine (SVM) classifier is trained to distinguish between the positive and negative sets for each invitee.
At test time, we are given a tracklet $T=\big\{\bm{t}_1,...,\bm{t}_N\big\}$ represented as a set of face feature vectors $\bm{t}_i\in{\mathbb{R}^d}$, and we classify each member $\bm{t}_i$ with the classifier of each identity $h$ and obtain a set of classification confidences $\big\{s\big(T\big)_{i,h}\big\}$. Hereinafter, we omit argument $T$ for brevity. We now aggregate the scores of each identity to obtain the final identity scores $s_h=\text{stat}\big(\big\{s_{i,h}\big\}_{i=1}^N\big)$
where $\text{stat}(\cdot)$ represents aggregation by e.g. taking the mean confidence.
When $s=\max_{h} s_h$ is smaller than a threshold, a new guest identity is added to 
$\mathcal{H}$, where the classifier for this person is trained by using $T$ as positive examples. 
$\{s_h\}_{h \in \mathcal{H}}$ is converted to a set of posterior probabilities $\{P(h | r, V)\}_{h \in \mathcal{H}}$ with a trained regression model. 

The adaptations we make over the original MBGS are as follows.
\begin{enumerate}
\item During SVM training we place a high weight over negative examples. The motivation here is to force training to classify regions of confusion as negatives e.g. if blurry positive and negative images get mapped to the same region in feature space we prefer to have negative confidence in this region.
\item We set $\text{stat}(\cdot)$ to be the function returning the $95^{\text{th}}$ percentile instead of the originally proposed mean function. The effect of this together with the previous bullet is that the final identity score is impacted by the most confident face instances in the tracklet and not the confusing ones, thereby mining the highest quality frames.
\item We augment an input feature vector with the cosine similarity score between the input and a face signature, which results in a classification function of the form of 
$\langle \bm{x},\bm{w}^h_{1:d} \rangle + w^h_{d+1}\cos\big(\bm{x}, \bm{q}_h\big)-b^h,$
where $\bm{x}\in{\mathbb{R}^d}$, $\bm{q}_h$ is $h$'s face signature obtained as the mean of the gallery face features of $h$, $\text{cos}(\cdot)$ is the cosine similarity, and $\big(\bm{w}^h,b^h\big)$ are linear weights and bias. We note that more complex rules tend to overfit due to the small size of enrollment, which typically consists of no more than 10 images.
\end{enumerate}

\section{Experimental Results}
\label{sec: experiments}
\vspace{-.5em}

We now report experimental results for the data described in Section \ref{sec: device}.
We first investigate certain aspects of the system by using the gold standard test set. 
Then, we show the results on the extended test set. 
The WERs were calculated with the NIST asclite tool. 
Speaker-attributed (SA) WERs were also calculated by 
scoring system outputs for individual speakers against the corresponding 
speakers' reference transcriptions\footnote{Note that SA-WER used here is different from SWER of \cite{FiscusEtAl:rt07}.}.

For speech recognition, we used a conventional hybrid system, consisting of a latency-controlled bidirectional long short-term memory (LSTM) acoustic  model (AM) \cite{Xue17} and a weighted finite state transducer decoder.
Our AM was trained on 33K hours of in-house audio data, including close-talking, distant-microphone, and artificially noise-corrupted speech. Decoding was performed with a 5-gram language model (LM) trained on 100B words. 
Whenever a silence segment longer than 300\,ms was detected, the decoder generated an n-best list, which was rescored with an LSTM-LM which consisted of two 2048-unit recurrent layers and was trained on 2B words.
To help calibrate the difficulty of the task, we note that the same models were used in our recent paper~\cite{Yoshioka19c}, where results on NIST RT-07 were shown.

The first row of Table \ref{tab: gold_standard_results} shows the proposed system's WERs for the gold standard test set. The WERs were calculated over all segments as well as those not containing overlapped periods. 
The second row shows the WERs of a conventional approach using single-output beamforming. Specifically, we replaced CSS in Fig.~\ref{fig: SRD} by a differential beamformer which was optimized for our device and ran speech recognition on the beamformed signal. In \cite{Boeddeker18}, we verified that our beamformer slightly outperformed a state-of-the-art mask-based MVDR beamformer.
The proposed system achieved a WER of 18.7\%, outperforming the system without CSS by 3.6 percentage points, or 16.1\% relative.
For single-speaker segments, the two systems yielded similar WERs, close to 15\%. From these results, we can see that CSS improved the recognition accuracy for overlapped segments, which accounted for about 50\% of all the segments. 


\begin{table}[t]
\centering
\caption{WERs on gold standard test set.}
\label{tab: gold_standard_results}
\vspace{.2em}
\begin{tabular}{|l|c|c|}
\hline
Front-end & All segments & No overlap \\ \Xhline{3\arrayrulewidth}
Single BF & 22.3 & 15.4  \\ 
CSS       & {\bf 18.7} & {\bf 15.1}  \\  \hline
\end{tabular}
\caption{SA-WERs on gold standard test set for different diarization configurations.}
\label{tab: SA-WER}
\vspace{.2em}
\begin{tabular}{|cc|c|c|}
\hline
\multirow{2}{*}{FaceID+SSL} & \multirow{2}{*}{SpeakerID} & \multicolumn{2}{c|}{Invited/Guest} \\
           &                       & 100\%/0\%  & 50\%/50\% \\ \Xhline{3\arrayrulewidth}
\checkmark &                       & 22.4 & 21.7 \\ 
\checkmark &            \checkmark & \textbf{19.8} & \textbf{20.4} \\ \hline
\end{tabular}
%
\caption{WER and SA-WER on extended test set.}
\label{tab: extended_results}
\vspace{.2em}
\begin{tabular}{|c|c|}
\hline
WER & SA-WER \\ \hline
20.1 & 22.1  \\ \hline
\end{tabular}
\vspace{-1em}
\end{table}

Table~\ref{tab: SA-WER} shows SA-WERs for two different diarization configurations and two different experiment setups. In the first setup, we assumed all attendees were invited to the meetings and therefore their face and voice signatures were available in advance. 
In the second setup, we used precomputed face and voice signatures for 50\% of the attendees and the other speakers were treated as `guests'. 
A diarization system using only face identification and SSL may be regarded as a baseline as this approach was widely used in previous audio-visual diarization studies~\cite{Hori12,Gebru18,Ban18}. 
The results show that the use of speaker identification substantially improved the speaker attribution accuracy. The SA-WERs were improved by 11.6\% and 6.0\% when the invited/guest ratios were 100/0 and 50/50, respectively. 
The small differences between the SA-WERs from Table \ref{tab: SA-WER} and the WER from Table \ref{tab: gold_standard_results} indicate very accurate speaker attribution. 

One noteworthy observation is that, if only face identification and SSL were used, a lower SA-WER was achieved when only 50\% of the attendees were known to the system. 
This was because matching incoming cropped face pictures against face snapshots taken separately under different conditions (invited speakers) tended to be more difficult than performing the matching against 
face images extracted from the same meeting (guest speakers).

Finally, Table \ref{tab: extended_results} shows the WER and SA-WER of the proposed system on the extended test set. 
For this experiment, we introduced approximations to the vision processing module to keep the real time factor smaller than one regardless of the number of faces detected. 
We can still observe similar WER and SA-WER numbers to those seen in the previous experiments, indicating the robustness of our proposed system.



\section{Conclusion}
\label{sec: conclusion}
\vspace{-.5em}

This paper described an online audio-visual meeting transcription system that can handle overlapped speech and achieve accurate diarization by combining multiple cues from different modalities. 
The SRD meeting transcription framework was proposed to take advantage of CSS. 
To the best of our knowledge, this is the first paper that demonstrated the benefit of speech separation in an end-to-end meeting transcription setting. 
As for diarization, a new audio-visual approach was proposed, which 
consumes the results of face tracking, face identification, SSL, and speaker identification as well as the TF masks generated by CSS for robust speaker attribution. 
Our improvements to face identification were also described. 
In addition to these technical contributions, 
we believe our results also helped clarify where the current technology stands.

\section{Acknowledgement}
\vspace{-.5em}

We thank Mike Emonts and Candace McKenna for data collection; Michael Zeng, Andreas Stolcke, and William Hinthorn for discussions; Microsoft Face Team for sharing their algorithms.

\vfill\pagebreak

\bibliographystyle{IEEEbib}
\bibliography{my_references,refs}

\end{document}